\documentclass{webofc}
\usepackage[varg]{txfonts}   

\begin{document}

\pagestyle{plain}

{\large 
 -- J-PARC HEF K1.8 BEAM LINE GROUP TECHNICAL REPORT  --
}

\section*
{Anti-deuteron beam study at J-PARC HEF K1.8 beam line
}

%
%

M.~Ukai$^{1,2}$, Y.~Ishikawa$^2$, T.~Takahashi$^1$,
K.~Tanida$^3$ and T.~O.~Yamamoto$^3$ 


~

{\small

\noindent
$^1$ {\it Institute of Particle and Nuclear Studies, High Energy Accelerator Research Organization (KEK), Tsukuba, Ibaraki 305-0801, Japan}

\noindent
$^2$ {\it Department of Physics, Tohoku University, Sendai, Miyagi 980-8578, Japan}

\noindent
$^3$ {\it Advanced Science Research Center, JAEA, Tokai, Ibaraki 319-1195, Japan}
}

\begin{center}
(Dataed: December 19, 2023)
\end{center}

\begin{center}
{\bf   Abstract}

~

\begin{minipage}[h]{11.5cm}

We performed a $\overline{d}$
beam study at the K1.8 beam line of 
J-PARC Hadron Experimental Facility.
1.8 GeV/$c$ $\overline{d}$ beam yield was measured
to be 0.30 $\pm$ 0.04 counts/spill for 
30 GeV  70 $\times 10^{12}$ protons/spill irradiated on a
66 mm thick of gold target
with
the vertical slit
opening widths of 2.2 mm, 5 mm and 5 mm for intermediate focus (IFV), mass slit 1 (MS1) and 2 (MS2), respectively.
Corresponding 
$\overline{p}$ beam yield is roughly estimated to be $\sim$ 0.3 Mcounts/spill for the same slit condition.
Then, the $\overline{d}/\overline{p}$ production ratio at extraction angle of 6 degrees
is estimated to be $\sim 10^{-6}$.
This is the first time measurement of  the $\overline{d}$ beam yield 
and $\overline{d}/\overline{p}$ production ratio at 
J-PARC.  
Further beam line tuning 
may increase the $\overline{d}$ beam yield.

\end{minipage}

\end{center}
%
%
\renewcommand{\arraystretch}{1.2}

\section{Overview}

The K1.8 beam line in the J-PARC Hadron Experimental Facility (HEF)
is a secondary beam line to deliver mass-separated secondary particles
at maximum momentum of 2 GeV/$c$
produced by a high-intensity proton beam from the Main Ring Synchrotron (MR).
Using meson beams such as $K^-$ and pions,
strangeness nuclear physics experiments are mainly conducted.
Use of other particle beams will extend our research fields and 
ability of  HEF.
$\overline{d}$  is one of the interesting beam particles.
Well controlled $\overline{d}$  beam is useful in 
particle and nuclear physics research.
%
However, production cross section data of $\overline{d}$  are insufficient
to estimate $\overline{d}$  beam intensity at secondary beam lines. 

We performed a 1.8 GeV/$c$ $\overline{d}$ beam study in May 2021
as the first trial of the $\overline{d}$ beam mode at the K1.8 beam line.
It was performed during the MR beam stability check period (3 hrs)
in a one-day interval of the E42 experiment running at the K1.8 beam line.



\section{Experimental setup}
\subsection*{K1.8 beam line and experimental area}
\begin{figure}[hb]
\includegraphics[width=11cm,clip]{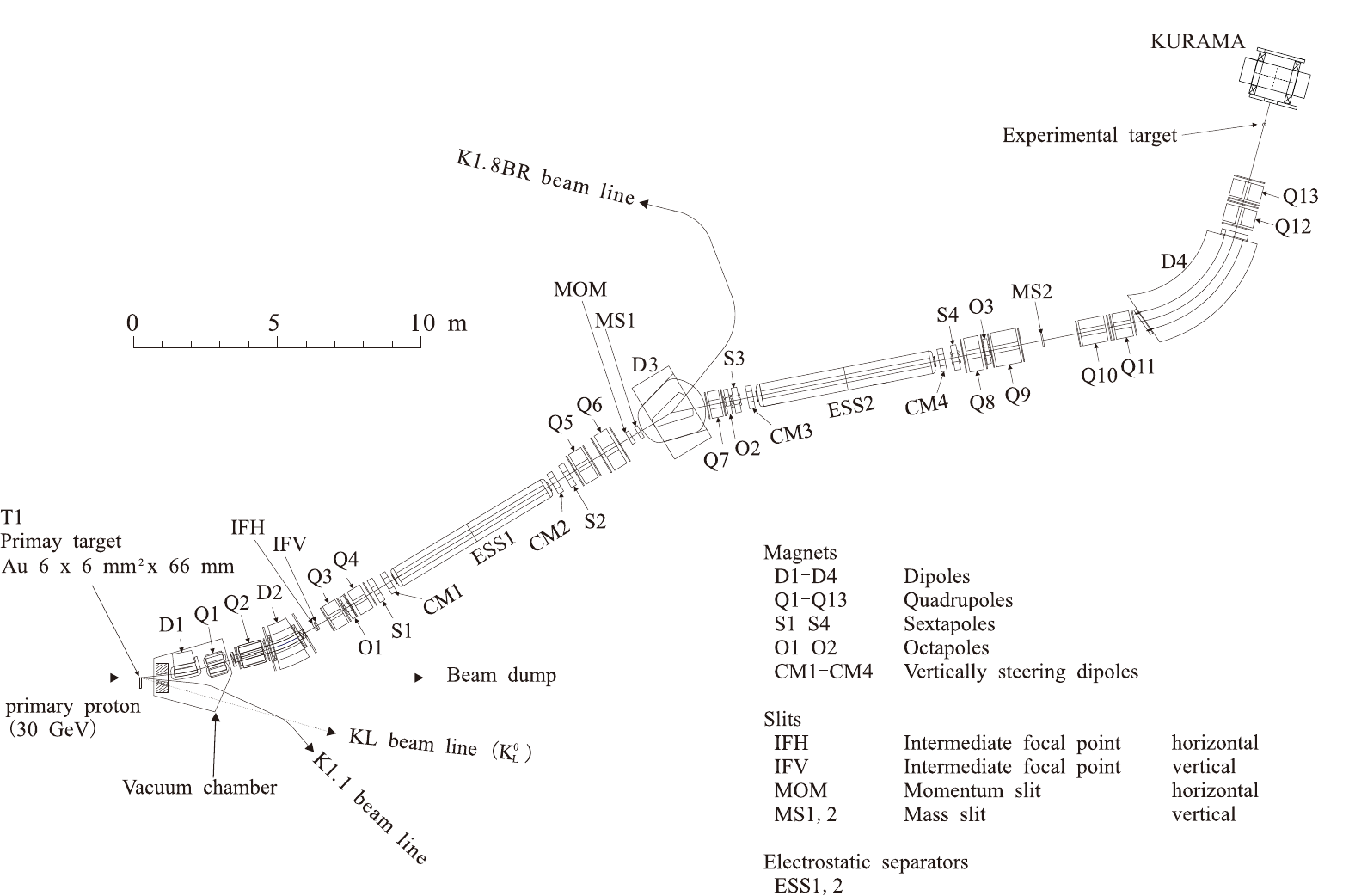}
\caption{
Schematic view of the K1.8 beam line.
}
\label{fig-k18bl}       
\end{figure}

The K1.8 beam line is a general-purpose mass-separated 
beam line that can supply secondary charged hadron beams 
of $K^{\pm}$, $\pi^{\pm}$, $p$ and $\overline{p}$ up to  2.0 GeV/$c$ \cite{PTEPk18}.
The layout of the K1.8 beam line elements is shown in Fig.~\ref{fig-k18bl}. 
Secondary particles are produced at a primary target (T1) 
by irradiating 30 GeV protons from the MR.
The current T1 target installed in 2019
is made of gold 
with a thickness of 66 mm, which corresponds to a 50\% beam loss.
%
%
In this study, 
the primary proton intensity on the T1 target 
was $70\times 10^{12}$ protons/spill. 
The primary proton beam conditions in the present study
are summarized in Table \ref{tab-cond}. 
The primary proton beam profile on the 
T1 target is also shown in Table \ref{tab-t1pro}.

\begin{table}[ht]
\centering
\caption{Primary proton beam conditions.
}
\label{tab-cond}       
\begin{tabular}{ccccccccc}
\hline
Acc.~run\# &  Proton energy & MR power & Protons/spill & Repetition & Spill length  \\
\hline
87 &30 GeV  & 64 kW      & 70$\times 10^{12}$   &5.2 s& 2.06 s  \\
\hline
\end{tabular}
\end{table}

\begin{table}[ht]
\centering
\caption{Typical primary proton beam profile at the T1 target.
}
\label{tab-t1pro}       
\begin{tabular}{ccccccc}
\hline
Mean(X) & Mean (Y) & Sigma(X) & Sigma (Y) \\ 
\hline
0.95 mm & $-$0.33 mm & 2.9 mm & 1.9 mm \\
\hline
\end{tabular}

\end{table}

\begin{table}[hb]
\centering
\caption{Specifications of the K1.8 beam line.} 
\label{tab-K18spec}       
\begin{tabular}{cccccccc}
\hline
Maximum momentum & 2.0 GeV/$c$ \\
Primary target (T1) &  Au 66-mm thickness \\
                                &  (from 2019 fall) \\
Extraction angle & 6 deg. \\
Momentum bite & $\pm $3\% \\
Beam line length & 46 m \\
\hline
\end{tabular}
\end{table}

Figure \ref{fig-k18bl} shows the schematic view of the K1.8 beam line.
Specifications of the 
K1.8 beam line and the T1 target are summarized in Table \ref{tab-K18spec}.
The extraction angle of the K1.8 beam line is 6 degrees. 
%
The beam line consists of 4 sectors; extraction part from the T1 to the intermediate focus (IF) point, two stages of electrostatic mass separators (from IF to mass slit 1 (MS1) and MS1 to MS2), and momentum analyzer.
%
%
At the intermediate focus point, the beam is vertically
focused in order to eliminate unwanted particles such as cloud pions from $K^0_S$ decay.
Each electrostatic separator (ESS) with  a 6 m long and 10 cm gap
is located between correction magnets (CM) which are vertical bending dipole magnets. 
In addition, mass slits (MS) are located downstream of CM2 and CM4.
Owing to these systems,
a high-purity kaon beam can be delivered to the K1.8 experimental area.

In this study, the beam momentum of 1.8 GeV/$c$ was chosen because it is 
well-studied and  the standard momentum at the K1.8 beam line.
IFV opening width was set to 2.2 mm ($^{+1.3}_{-0.9}$ mm)  
to keep the same 
value as the E42 condition.
%
ESS1 was set to $\pm$150 kV and ESS2 was OFF,
while both ESS1 and ESS2 were operated with $\pm$250 kV in the E42 run for $K^-$ beam.
Then, currents of CM1 and CM2 were tuned for corresponding values for $\overline{p}$ 
and $\overline{d}$, and CM3 and CM4 were OFF.
CM1 and CM2 currents were set to the same value without offset setting.
Currents of K1.8 beam line magnets,  except for CMs,  
were set to the same values as the E42 $K^-$ beam settings (1.8 GeV/$c$) 
optimized to the ESS1/ESS2 = $\pm$250 kV operation.

\begin{table}[hb]
\centering
\caption{K1.8 beam line and spectrometer conditions.
}
\label{tab-K18}       
\begin{tabular}{cccccccc}
\hline
K1.8D1 current &momentum  & beam polarity & ESS1    & ESS2& KURAMA \\ 
\hline
$-721$ A  & 1.8 GeV/$c$  &  negative & $\pm$150 kV  & OFF  & OFF \\
\hline
\end{tabular}
\end{table}

\subsection*{K1.8 detectors}

\begin{figure}[h]
\centering
\includegraphics[width=8cm,clip]{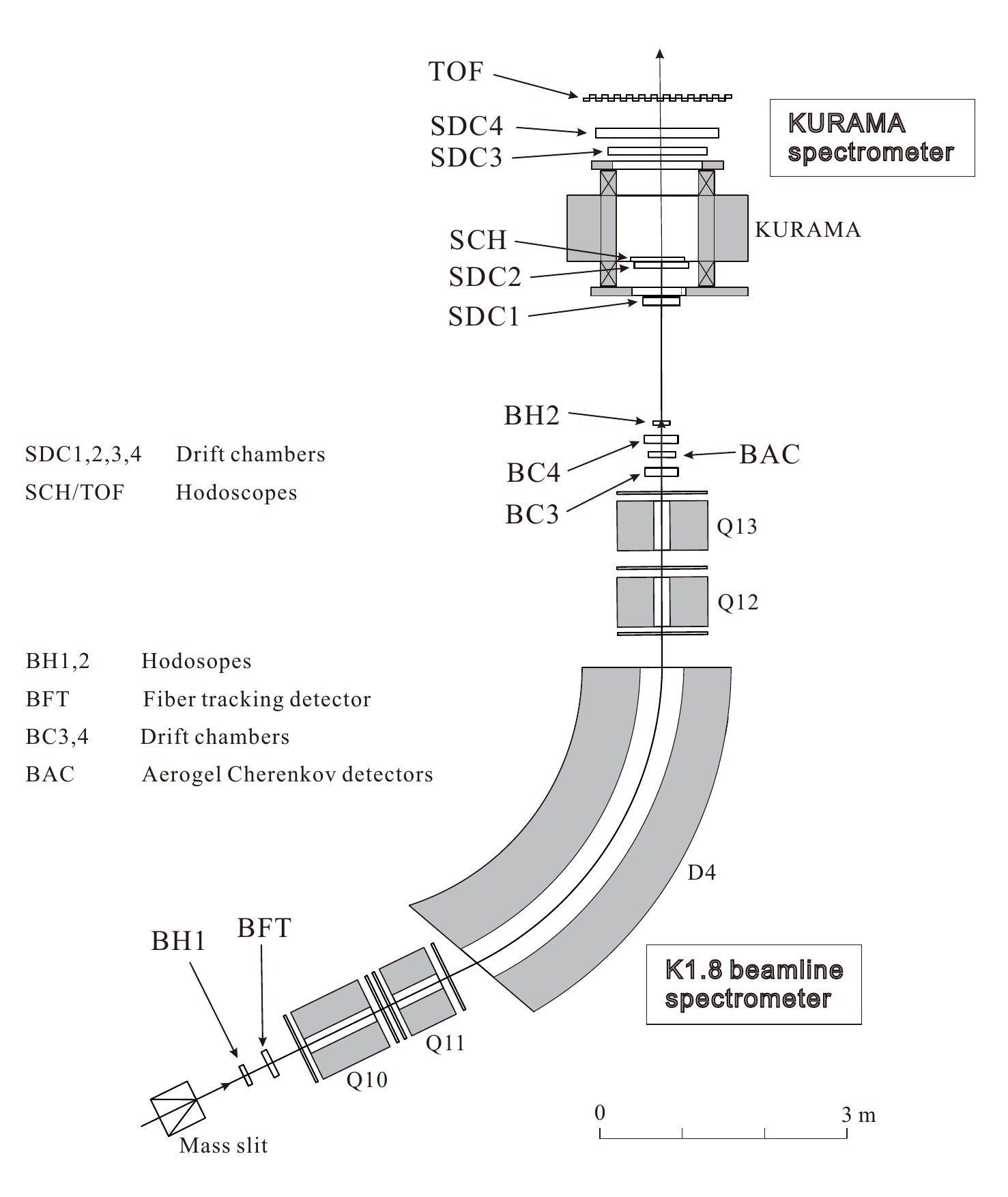}
\caption{Schematic view of the K1.8 experimental area.
KURAMA is a dipole magnet to analyze the scattered particle momenta for reaction spectroscopy.
In this study, KURAMA was off.
Some detectors of the E42 experiment are omitted here. 
}
\label{fig-KURAMA}  
\end{figure}

Figure \ref{fig-KURAMA} shows the schematic view of the
K1.8 experimental area.
Beam particles were identified by threshold-type aerogel Cherenkov counter (BAC) 
in the on-line trigger and Time-of-Flight counters (BH1, BH2 and TOF)
in the off-line analysis. 
BH2 is the time-zero counter for all detectors.
Specifications of these counters are summrized in Table \ref{tab-area}. 
Path lengths between BH1-BH2 (BTOF) and  BH2-TOF (STOF)
are 10.2 m and  4.1 m, respectively.
BFT (plastic scintillation fiber tracker) and BC3, 4 (drift chamber) were also used
to check the beam momentum and profile.

\begin{table}[h]
\centering
\caption{Specifications of plastic scintillator hodoscopes (BH1, BH2 and TOF)
and $n$=1.03 aerogel Cherenkov counter (BAC).}
\label{tab-area}       
\begin{tabular}{ccccc|ccc}
\hline
 Detector & Effective area & Segmentation &  Path length \\
               &  W $\times$ H $\times$ T [mm$^3$] &     & from BH2 [m] \\
 
\hline
BH1 & $170 \times 66 \times 5$ & 11 &  $-$10.2 (BTOF length)\\
BAC & $342 \times 80 \times 66 $ & 1 & $-0.4$ \\                                   
BH2 &  $118  \times 80 \times 5$         &   8  &  -- \\
TOF & $1805 \times 1800 \times 30 $   & 24   & 4.1(STOF length) \\
    
\hline
\end{tabular}
\end{table}

Table \ref{tab-tof} shows the velocities, relative energy deposit $dE/dx$ and 
corresponding  BTOF/STOF timing of 
each beam particle for measured average momentum 
of 1.82 GeV/$c$ with K1.8D1 = $-$721 A setting.
The relative energy deposit ($dE/dx$) values are calculated by 
the Bethe-Bloch equation
normalized to that of the $\pi^-$.
$\pi$ timing is calibrated to be 0 ns in the BTOF/STOF histograms.
Then, time differences from $\pi$ (Tdiff) are also shown.
In addition, 
time differences caused 
by the measured momentum dispersion of 
 $\delta p \sim \pm 5$ \% are also shown.

\begin{table}[htb]
\centering
\caption{Velocities, relative energy deposit ($dE/dx$) 
and  BTOF/STOF time difference from $\pi$ of 1.82 GeV/$c$ beam particles.
 $dE/dx$ is the relative value to $\pi$ calculated by the Bethe-Bloch equation
(arbitrary unit).  
}
\label{tab-tof}       
\begin{tabular}{cccc|ccr|ccc}
\hline 
         
      & mass &    $\beta$ & $dE/dx$
& {\small BTOF} & {\small $\Delta T$($\delta p \pm$5\%)} &
 Tdiff& {\small STOF}  & Tdiff \\
        &[GeV/$c^2$] &     & [a.u.] & [ns] & [ns] & [ns]  & [ns] & [ns] \\
\hline

$\pi$ &  0.1396 &  0.9971 & 1 &  34.124  & $^{+0.009}_{-0.011}$ & -- ~~~ 
& 13.716 & --   \\

$K$   &   0.4937 & 0.9651 & 0.93 &  35.253    & $^{+0.112}_{-0.130}$& 1.230
& 14.170 &  0.454 \\

$p$ &     0.9383 & 0.8888  & 1.01 &38.279   & $^{+0.375}_{-0.431}$ & 4.156
&15.387 &  1.670 \\

$d$ &    1.8757 &  0.6964  & 1.49 &48.858  &$^{+1.184}_{-1.340}$ & 14.735 
&19.639 &  5.923 \\

\hline

\end{tabular}
\end{table}

\subsection*{Run summary}

In the present study,
data for three CM current conditions corresponding to 
$\overline{d}$, $\overline{p}$ and $\pi^-$ beam settings were acquired.
$\overline{p}$ and $\pi^-$ beam data were taken for time calibration of 
BTOF and STOF distributions.
Table \ref{tab-runsum} shows the 
run summary.
Slit conditions in the  $\overline{d}$ beam run (RUN\#5388) and $\overline{p}$ 
beam run (RUN\#5390)
were the same except for MS1, 2. 
IFH, MS1 and MS2 widths of 
$\pi^-$ beam run (RUN\#5392)  were $\pm$30 mm, $\pm$0.5 mm and $\pm$0.45 mm,
respectively.
IFV condition was the same.

\begin{table}[h]
\centering
\caption{RUN summary. 
}
\label{tab-runsum}       
\begin{tabular}{ccccccccc}
\hline
RUN \# & Run setting & CM1, 2 & MS1, 2 &
Trigger  &Prescale & DAQ eff.
&  Spill  \\
\hline 
5388   & $\overline{d}$ beam & 332 A & $\pm$ 2.5 mm
& BH2 $\times \overline{\rm BAC} $ & 1/2 & 98\%
& 326 \\
5390   & $\overline{p}$ beam & 259 A& $\pm$ 0.7 mm
& BH2 $\times \overline{\rm BAC} $& & 
&  81 \\
5392   & $\pi^-$ beam & 227 A &   
& BH2 & & 
& 79 \\

\hline
\end{tabular}
\end{table}

\clearpage

\section{Anti-deuteron beam study}
\label{sec-dbar}

Since CM current value is basically proportional to $1/\beta$ of beam particles,
332 A was set for CM1 and CM2 as the $\overline{d}$ beam setting estimated 
from 259 A which was the optimized value for the $\overline{p}$ beam
setting as described in Sec.~\ref{sec-pbar}.
Though the best CM currents may be slightly off from the estimated one using $1/\beta$, 
CM1, 2 currents were not scanned
but fixed due to the limit of the study time.
In addition, from the previous K1.8 beam line study, 
the offset value of  CM current was known to be non-zero to maximize the beam yield 
but was omitted here. The offset value of CM current  is described in Appendix B.
Slits, ESSs and CMs conditions for this study are summarized in Table \ref{tab-dcond}.

\begin{table}[h]
\centering
\caption{
Slit, ESS and CM conditions in the $\overline{d}$ beam run (RUN\#5388).
}
\label{tab-dcond}       
\begin{tabular}{cccccccccc}
\hline
IFV  & IFH & MOM & MS1/MS2   & ESS1 & ESS2 & CM1/CM2 & CM3/CM4 \\
$[$mm$]$ & $[$mm$]$ & $[$mm$]$  
&  $[$mm$]$ & $[$kV$]$ & $[$kV$]$ & $[$A$]$ & $[$A$]$\\
\hline
$^{+1.3}_{-0.9}$  & $\pm$120 & $\pm$180  & $\pm$2.5/$\pm$2.5 & $\pm $150 &  0  & 332 & 0  \\  
\hline
\end{tabular}

\end{table}

BH2 $\times \overline{\rm BAC}$ trigger data were acquired for $\sim$30 min (326 spills).
BH2 $\times \overline{\rm BAC}$ rate was 4.2 kcounts/spill.
To keep DAQ efficiency at $\sim$ 100\%,
1/2 prescale was applied to the trigger.
Counting rates of each counter  are summarized in Table \ref{tab-5388rate}.

\begin{table}[hb]
\centering
\caption{Counting rates (counts/spill) for RUN\#5388 ($\overline{d}$ beam run). 
}
\label{tab-5388rate}       
\begin{tabular}{cccccccc}
\hline
BH1  & BH2 & BAC & TOF  & BH2 $\times \overline{\rm BAC}$ \\ 
\hline
  607k  & 367k &443k & 511k  & 4.2k  \\
\hline
\end{tabular}
\end{table}

Due to the large accidental background,
no clear peak around the $\overline{d}$ timing
was observed in both BTOF and STOF time distribution for the 
BH2 $\times \overline{\rm BAC}$ trigger data.
Then,  $\overline{d}$ timing cuts were applied as follows,

\begin{itemize}
\item Upstream $\overline{d}$ cut:  $-16.7$ < BTOF < $-12.7$ ns \& number of BH1 cluster  = 1,
\item Downstream $\overline{d}$ cut: 3.9 < STOF  < 7.9  ns \& TOF hit segment 8--15.
\end{itemize}

\noindent
The time gate of $\pm$2 ns was determined from  
the $\pi^-$ peak widths of 0.2 ns($\sigma$)  for BTOF and STOF and 
measured momentum distribution ($\sim \pm 5$ \%) 
as shown in Table \ref{tab-tof}. 

\begin{figure}[ht]
\includegraphics[width=13cm,clip]{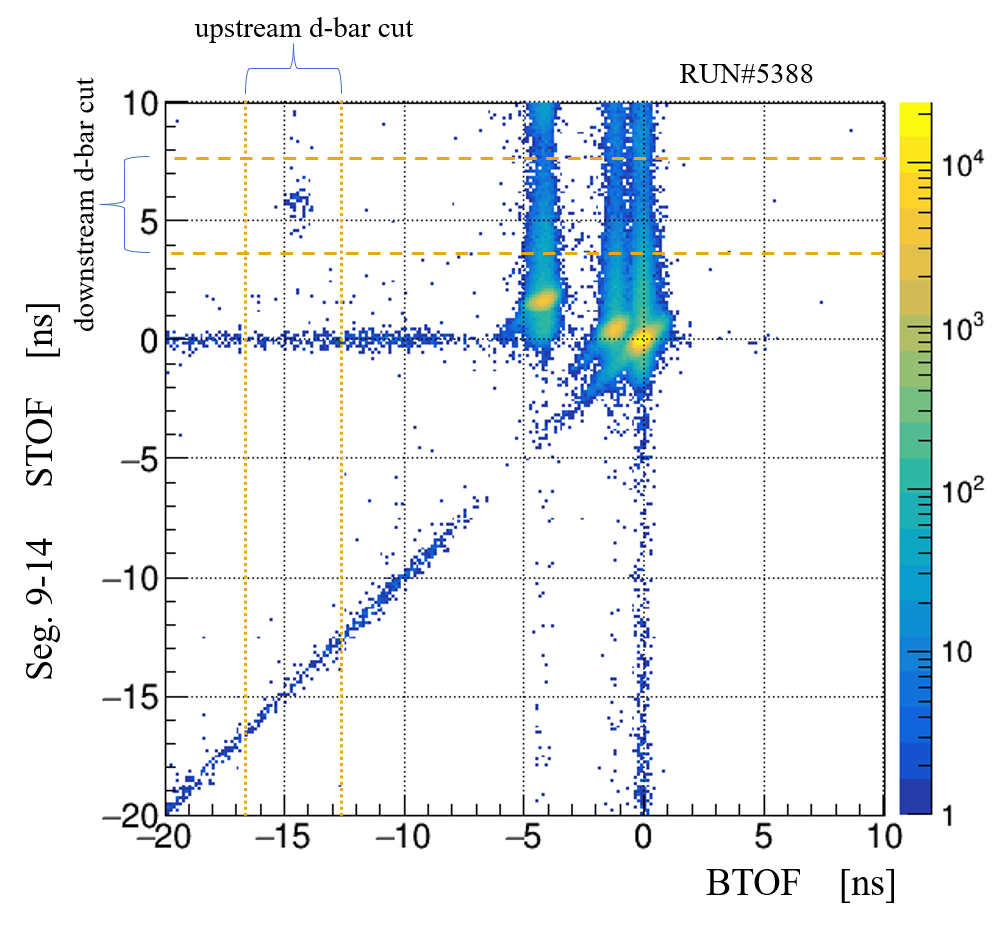}
\caption{Correlation between BTOF (BH1-BH2) and STOF (BH2-TOF)
for BH2 $\times \overline{\rm BAC}$ trigger data (RUN\# 5388).
Timing gates for the upstream $\overline{d}$ cut (dotted lines) and the
downstream   $\overline{d}$ cut (dashed lines) are also shown.
}
\label{fig-bstof}  
\end{figure}

\begin{figure}[ht]
\includegraphics[width=13cm,clip]{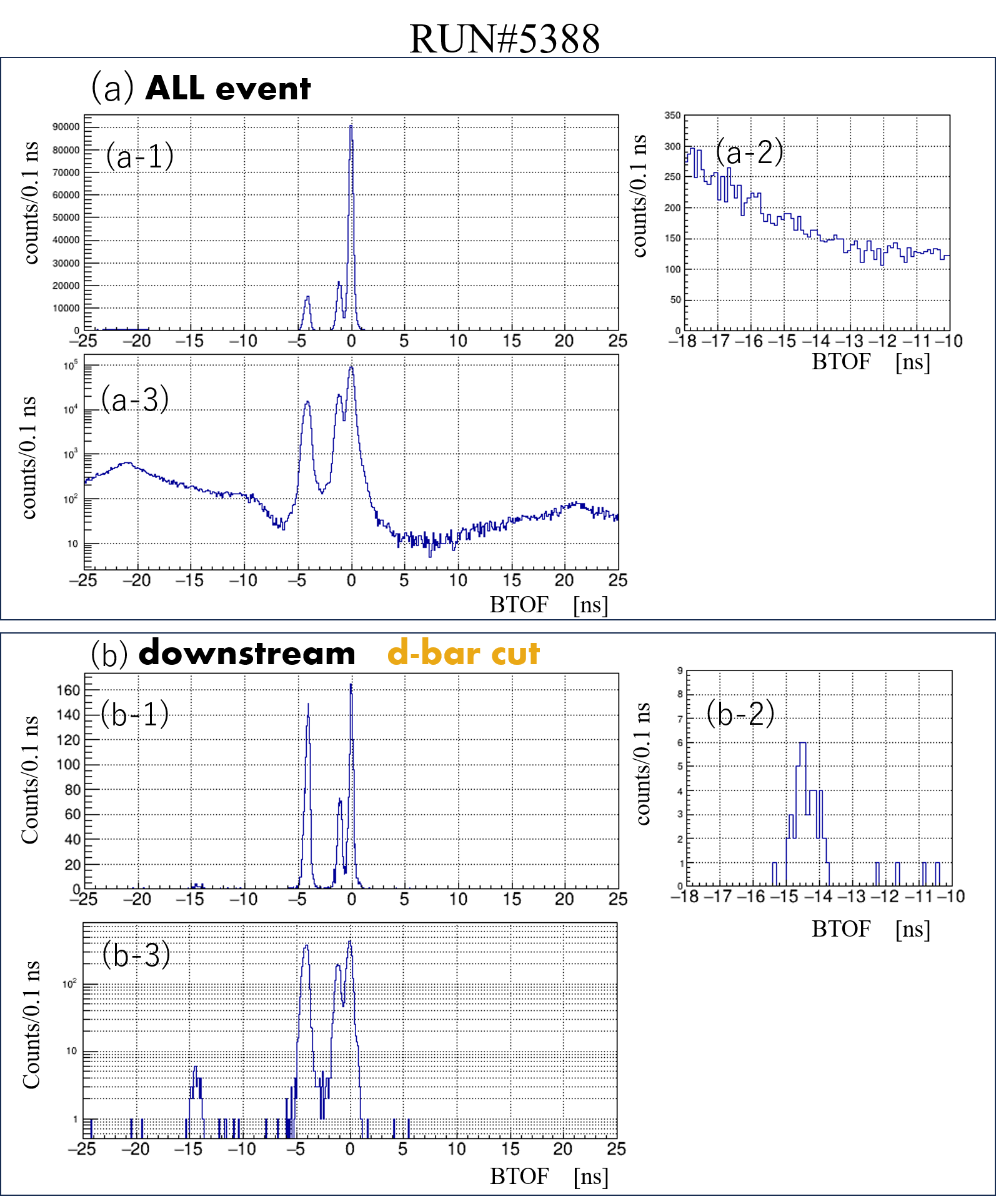}
\caption{BTOF (BH1-BH2) time distributions for 
BH2 $\times \overline{\rm BAC}$ trigger data (RUN\# 5388).
(a) is for all events. 
(b) is for STOF timing to be $\overline{d}$ timing and TOF hit segment to be 8--15 
(downstream $\overline{d}$ cut).
(a-2) and (b-2) are enlarged view around $\overline{d}$ timing and 
(a-3) and (b-3) are log scale 
of (a-1) and (b-1), respectively.  
In the case of the coincidental hits of staggered segments, only one of the pairs is included
in (b).
a 45-counts peak structure was found at BTOF $\overline{d}$ timing ($-$14.7 ns) in (b).
}
\label{fig-btof}  
\end{figure}

\begin{figure}[hb]
\includegraphics[width=13cm,clip]{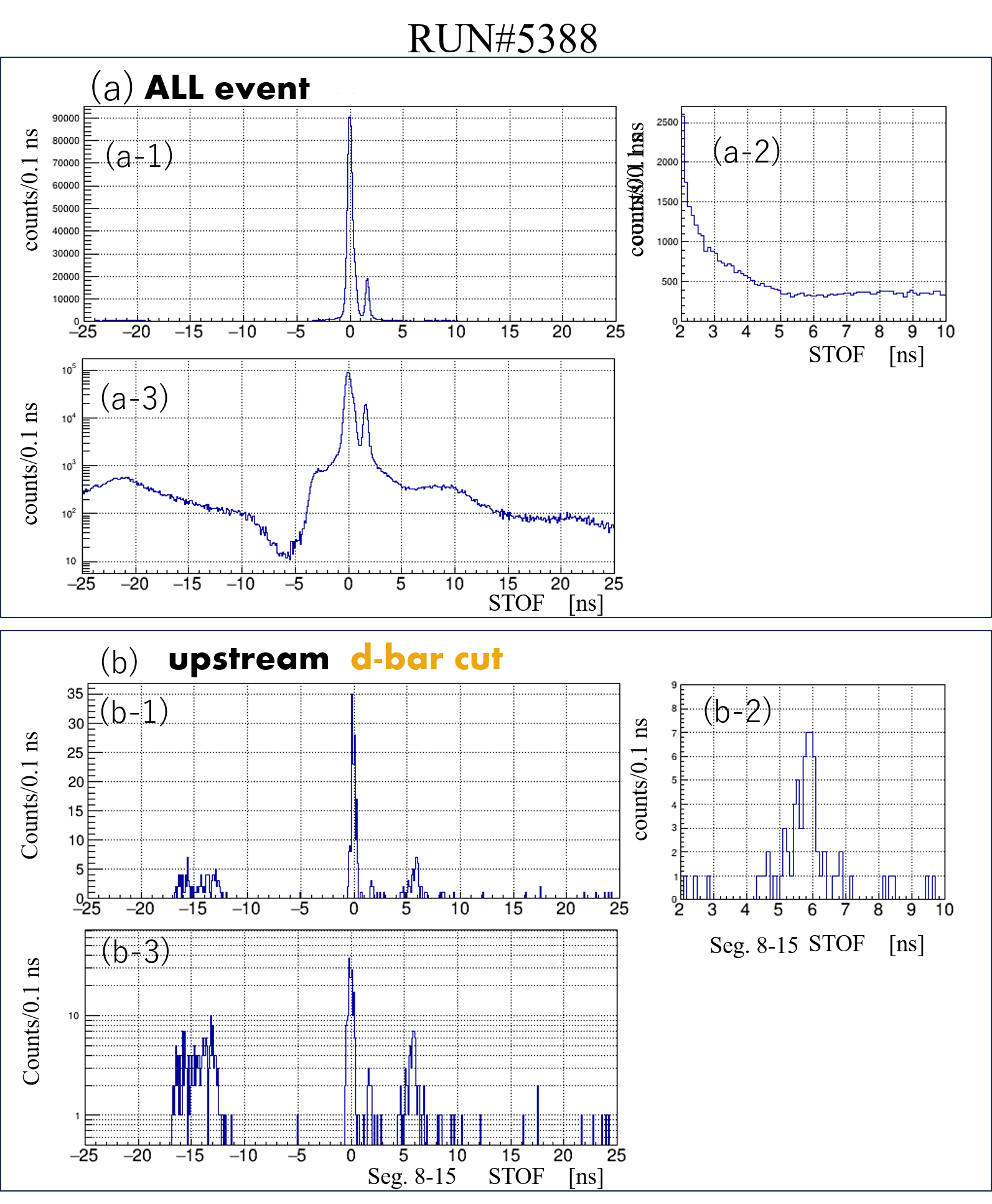}
\caption{STOF (BH2-TOF) time distributions 
for BH2 $\times \overline{\rm BAC}$ trigger
 data (RUN\# 5388).
(a) is for all events. 
(b) is STOF with TOF segment to be 8 -- 15 
for events with BTOF timing to be $\overline{d}$ timing (upstream $\overline{d}$ cut).
(a-2) and (b-2) are enlarged view around $\overline{d}$ timing 
and (a-3) and (b-3) are log scale 
of (a-1) 
and (b-1), respectively.
Both of coincidental hits of staggered TOF segments are included in histograms.  
}
\label{fig-stof}  
\end{figure}

Figure \ref{fig-bstof} show the correlation between BTOF and STOF.
Timing gates for the upstream $\overline{d}$ cut (dotted lines) and 
downstream   $\overline{d}$ cut (dashed lines) are also shown.
Figure \ref{fig-btof} shows the BTOF distribution for all events in (a) , 
and those with downstream $\overline{d}$ cut in (b).
In Fig.~\ref{fig-btof}(b), 
in the case of the coincidental hits of staggered segments, only one of the pairs is included in the histogram.
As shown in Fig.~\ref{fig-btof}(b), 
a peak structure at the  $\overline{d}$ timing was found with 45 counts.
%
In addition, Fig.~\ref{fig-stof} shows the STOF distribution 
for all event in (a), 
and for events with upstream  $\overline{d}$  cut and TOF segment to be 8 -- 15 in (b).
In Fig.~\ref{fig-stof}(b), 
coincidental hits of staggered TOF segments are included both of the pair hits.
For both Figs.~\ref{fig-btof} and \ref{fig-stof} ,
(a-2) and (b-2) are enlarged histogram around the $\overline{d}$ timing 
and (a-3) and (b-3) are log  scale
of (a-1) and (b-1), respectively.
In Fig.~\ref{fig-stof}(b), structure around $-$15 ns 
in the STOF distribution corresponds to events of accidental $\pi^-$ beam enhanced by 
the BTOF cut of $-16.7$ --  $-12.7$ ns.

Taking into account the trigger prescale factor (0.5), DAQ efficiency (0.98) 
and BH1 multiplicity cut efficiency (0.93),
the $\overline{d}$ beam yield for 361 spills was obtained to be 99 $\pm$ 15 counts 
using the result of Fig.~\ref{fig-btof}(d).
Then, the $\overline{d}$ beam yield at the K1.8 beam line 
is obtained to be 0.30 $\pm$ 0.04 counts/spill
for the primary beam of  30 GeV $70 \times 10^{12}$ protons/spill 
irradiated on the 66-mm thick Au target.
Since the beam line magnets were not well tuned for this ESS condition
as described in Appendix B,
a higher $\overline{d}$ beam yield is expected with further study.


\clearpage

\clearpage

\begin{figure}[h]
\centering
\includegraphics[width=7cm,clip]{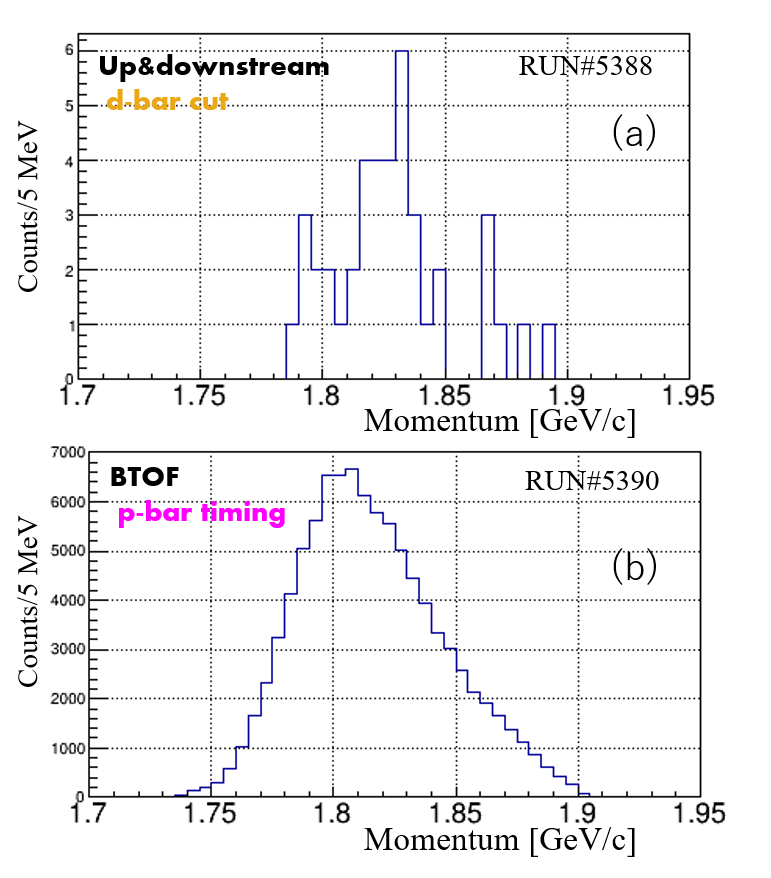}
\caption{
Momentum distribution for (a) $\overline{d}$ beam and 
(b) $\overline{p}$ beam timing events.
}
\label{fig-pk18pd}  

\centering
\includegraphics[width=7cm,clip]{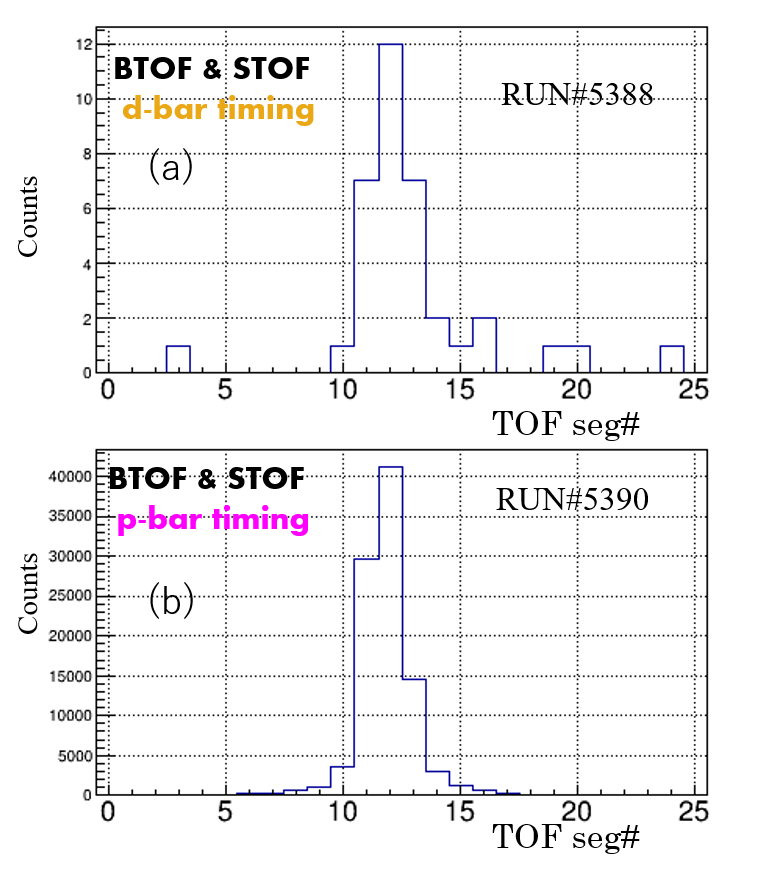}
\caption{Hit pattern of  TOF counter for (a) $\overline{d}$ beam and (b) $\overline{p}$ beam timing events.
}
\label{fig-tofseg}       
\end{figure}

Figures \ref{fig-pk18pd} and \ref{fig-tofseg}
show the momentum distributions and hit pattern of TOF counter for
(a) $\overline{d}$ beam and (b) $\overline{p}$ beam timing events 
taken from 
RUN\#5388 and RUN\#5390, respectively.

\clearpage

\begin{figure}[h]
\includegraphics[width=10cm,clip]{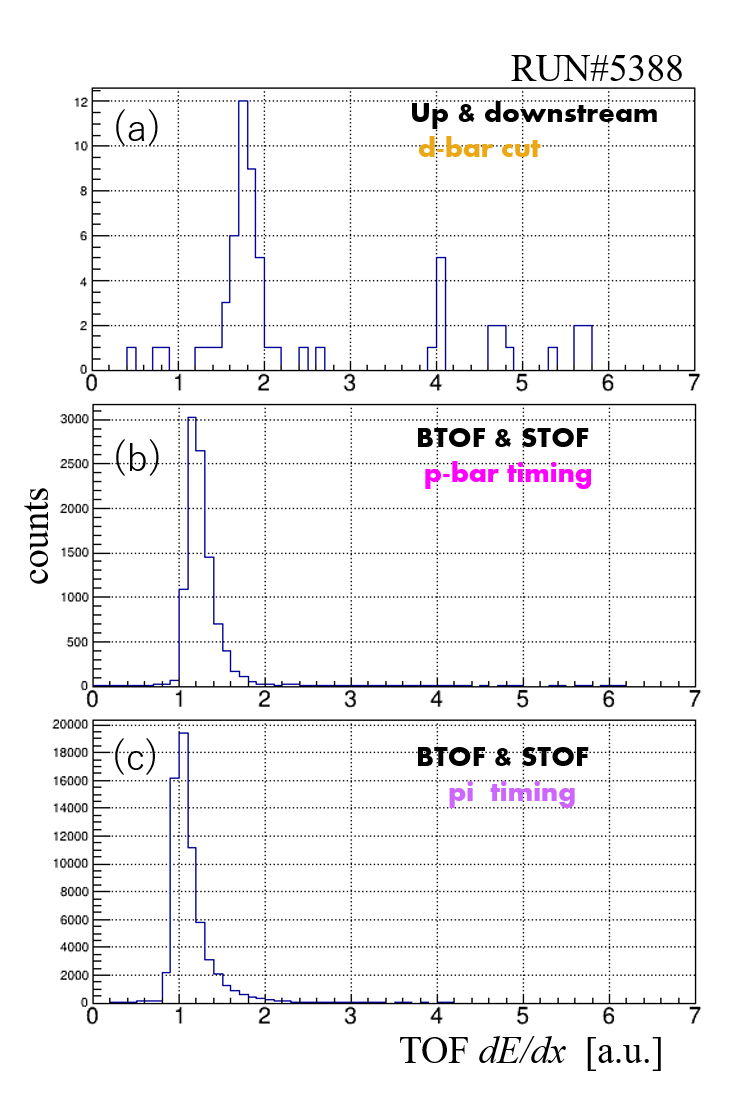}
\caption{Relative energy deposit ($dE/dx$) of TOF counter (segment \# 8 -- 15).  
The peak position for $\pi ^-$ beam in (c) is calibrated to be 1.
}
\label{fig-tofde}       
\end{figure}

Figure \ref{fig-tofde}
shows the relative energy deposit ($dE/dx$) spectra of TOF counter (segment \# 8 -- 15)
for (a) $\overline{d}$, (b) $\overline{p}$ and (c) $\pi ^-$ timing events 
selected using BTOF and STOF information.
In these spectra, the  $dE/dx$ peak for $\pi^-$ beam (RUN\# 5392)
is calibrated to be 1.
Fit results of these 
peak positions are 
1.8 $\pm$ 0.2, 1.2 $\pm$ 0.1 and 1.0 $\pm$ 0.1, respectively.
The measured $dE/dx$ ratio between (a) and (b) of 1.5 
confirms that 
the observed events are identified as $\overline{d}$ particles.

\clearpage

\section{Anti-proton beam study and expected yield}
\label{sec-pbar}

Prior to the $\overline{d}$ beam data acquisition,
$\overline{p}$ beam tuning was performed with narrow slit conditions 
to optimize the CM current for $\overline{p}$ beam using the visual scaler.
After the $\overline{d}$ beam data was acquired, 
the CM current settings were return to the  
$\overline{p}$ beam setting, then its data were acquired.
Slits, ESS and CM conditions are summarized in Table \ref{tab-pcond}.

\begin{table}[h]
\centering
\caption{Slit, ESS and CM conditions in the  $\overline{p}$ beam run (RUN\#5390).
}
\label{tab-pcond}       
\begin{tabular}{cccccccccc}
\hline
IFV  & IFH & MOM & MS1/MS2   & ESS1 & ESS2 & CM1/CM2 & CM3/CM4 \\
$[$mm$]$ & $[$mm$]$ & $[$mm$]$  
&  $[$mm$]$ & $[$kV$]$ & $[$kV$]$ & $[$A$]$ & $[$A$]$\\
\hline
$^{+1.3}_{-0.9}$  & $\pm$120 & $\pm$180  & $\pm$0.7/$\pm$0.7 & $\pm $150 &  0  & 259 & 0  \\  
\hline
\end{tabular}

\end{table}

To optimize beam line parameters
for the $\overline{p}$ beam setting,
CM1 and CM2 currents 
were scanned to maximize the scaler counts of 
BH2 $\times \overline {\rm BAC} $ as $\overline{p}$ beam yield.
Then, the maximum $\overline{p}$ beam yield was obtained 
with CM1, 2 = 259 A.
After the $\overline{p}$ beam tuning, MS1, 2  and CM1, 2 current settings were 
changed to $\overline{d}$ beam mode, and after that, 
MS1, 2  and CM1, 2 setting were returned to $\overline{p}$ beam setting to 
take the BH2 $\times \overline {\rm BAC} $ trigger data as the $\overline{p}$ beam setting.
%
The counting rates of counters for RUN\#5390
are summarized 
in Table \ref{tab-pbar}.
From the BTOF distribution, 
95\% of BH2 $\times \overline{\rm BAC}$  triggers were found to be  
$\overline{p}$ beam.
Then, the $\overline{p}$ beam yield for MS1, 2 = $\pm$0.7 mm was estimated to be 
95 kcounts/spill.
$\overline{p}$ beam yield ratio  between MS1, 2 = $\pm$2.5 mm and MS1, 2 = $0.7$ mm is 
roughly estimated to be 2.6 $\sim$ 3.0 as described in  Appendix A.
The $\overline{p}$ beam yield for MS1, 2 = $\pm$2.5 mm was 
roughly estimated to be $\sim$ 0.3 Mcounts/spill.

\begin{table}[hb]
\centering
\caption{
Counting rates (counts/spill) of counters  for RUN \#5390 ($\overline{p}$ beam).
}
\label{tab-pbar}       
\begin{tabular}{ccccccc}
\hline
BH1 & BH2   &  BAC  & TOF  &  
BH2 $\times \overline {\rm BAC} $  
\\   
\hline
796k & 377k & 377k & 564k   & 100k 
\\
\hline

\end{tabular}
\end{table}


\section{Anti-deuteron/anti-proton production ratio}

The $\overline{d}$ beam yield was 
obtained to be 0.30 $\pm$ 0.04 counts/spill as described 
in Sec.~\ref{sec-dbar}.
Corresponding  $\overline{p}$ beam yield was roughly estimated to be 0.3 Mcounts/spill.
Then, the $\overline{d}$/$\overline{p}$ production ratio  
for  30 GeV proton irradiated on an Au target at 6 degree extraction angle
was obtained to be the order of $\sim 10^{-6}$.


\section{Prospect of anti-deuteron beam yield}

In this study, the $\overline{p}$ 
beam yield was estimated to be $\sim$ 0.3 Mcounts/spill for 
ESS1 = $\pm$150 kV and ESS2 = OFF with MS1, 2 = $\pm$2.5 mm.
However, assuming the $K^-$/$\overline{p}$ 
ratio to be 0.7 $\sim$ 1 as described in  Appendix B,
$\overline{p}$ beam yield  is expected to be 0.7 $\sim$ 1 Mcounts/spill
for ESS1, 2 = $\pm$250 kV with MS1, 2 = $\pm$2.5 mm.
This is  because that the beam line magnet parameters were not optimum for
the present study
but for ESS1, 2 = $\pm$250 kV with non-zero CM offset condition. 
Then, the $\overline{d}$ beam yield is expected to also increase by 
2.3 $\sim$ 3.3 times.
In addition, the IFV width of 2.2 mm was narrower than the 
primary proton beam profile at the T1 target of 1.9 mm ($\sigma$). 
If the vertical profile on the IF point is 1.9 mm($\sigma$), 
only 43\% of the  beam can pass through the IFV width of  2.2 mm.
Then, the yield is expected to increase by $\sim$2 times for wider IFV width.
However, such wide IFV width causes a huge background.
To get more $\overline{d}$ beam intensity with lower background contamination,
detailed beam line tuning is necessary.
In addition, to estimate a realistic gain factor of $\overline{d}$ beam yield,  
the beam line simulation using such as TURTLE code should be performed.  

\section{Summary}

We performed the $\overline{d}$ beam study at the K1.8 beam line.
As the result,
1.8 GeV/$c$ $\overline{d}$ beam yield was measured 
to be 0.30 $\pm$ 0.04 counts/spill for 
30 GeV  70 $\times 10^{12}$ protons/spill irradiated on a
66 mm thick Au target at the 
the extraction angle of 6 degrees.
In this study, the vertical slit 
(IFV,  MS1 and  MS2) opening widths were set to 2.2 mm, 5 mm and 5 mm, respectively.
Corresponding 
$\overline{p}$ beam yield is 
roughly estimated to be $\sim$0.3 Mcounts/spill for the same slit condition.
Then, the $\overline{d}$/$\overline{p}$ production ratio at 6 degrees
is estimated to be $\sim 10^{-6}$.
This is the first time we have  measured the $\overline{d}$ beam yield and 
$\overline{d}$/$\overline{p}$ production ratio at 
J-PARC.  
Further beam line tuning 
may increase the $\overline{d}$ beam yield.





\section*{Appendix A ~~Beam yield vs MS 1, 2 opening width} 

\begin{figure}[h]
\includegraphics[width=11cm,clip]{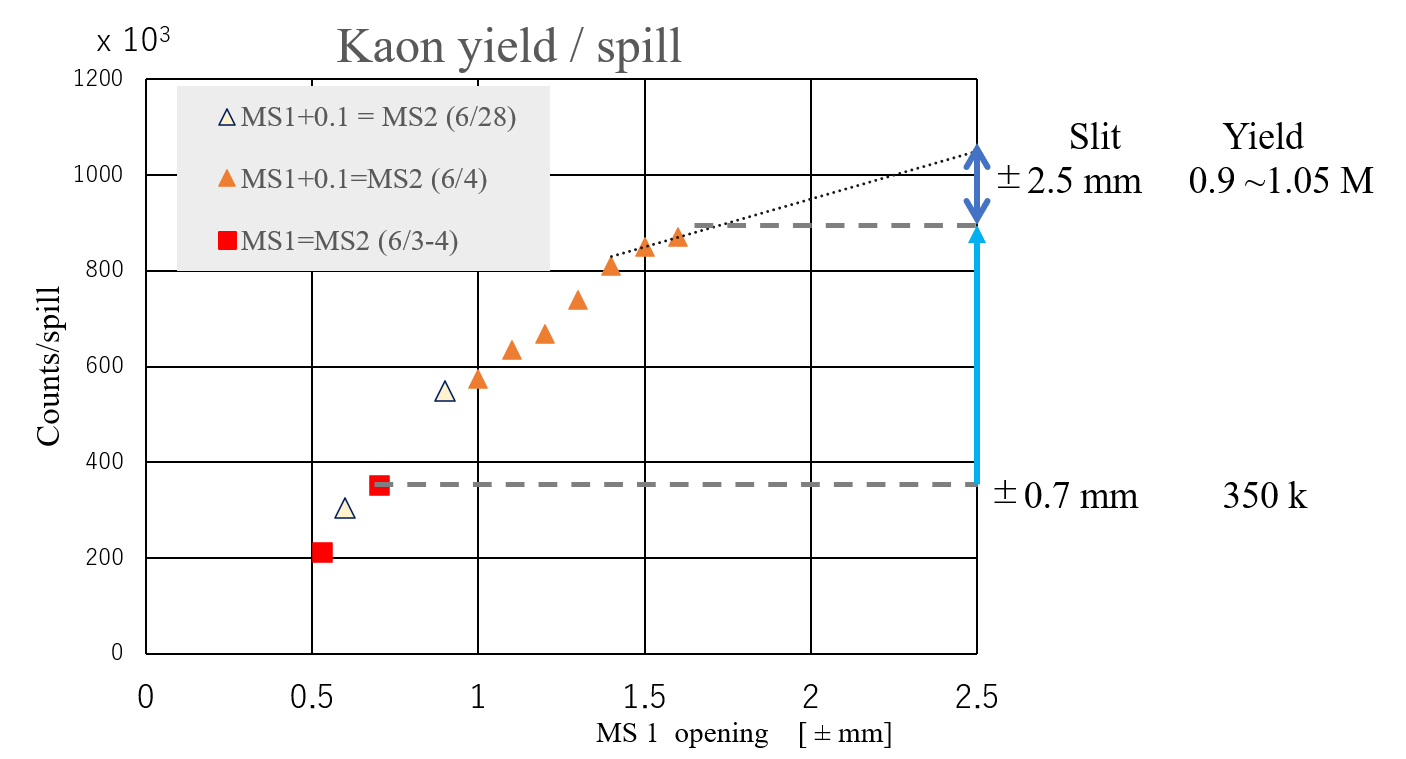}
\caption{1.8 GeV/$c$
$K^-$ beam yield as a function of the MS1 opening for 
IFV = $^{+1.3}_{-0.9}$ mm and ESS1, 2 = $\pm$250 kV with 
CM offset measured 
in Acc.~RUN87 (2021 June) with $70 \times 10^{12}$ protons/spill on the T1 target.
Beam time conditions are summarized in Table \ref{tab-kcond}.
MS2 openings almost follow MS1 openings.
}
\label{fig-kvsslit}  
\end{figure}

Mass slit width dependence of 1.8 GeV/$c$ $K^-$ beam yield was measured
for the E42 physics run in the same MR cycle with 
the same MR power (70 $\times 10^{12}$ protons/spill). 
The results are summarized in Fig.~\ref{fig-kvsslit}.
The Slit, ESS and CM conditions are summrized in Table \ref{tab-kcond}.
%
Since beam line magnet conditions were well
tuned  for physic data taking of the E42 experiment,
the result shows  the best values of $K^-$ beam yield. 

As shown in Fig.~\ref{fig-kvsslit},
the $K^-$ beam yield seems to be saturating at MS1 width of $\pm$ 1.5 mm
for  IFV width of 2.2 mm.
By assuming linear extrapolation from MS1 = $\pm$1.5mm and $\pm$1.6 mm yield,
1.05 Mcounts/spill  for MS1 = $\pm$2.5 mm was obtained at the maximum.
Then, the $K^-$ yield is estimated to be 0.9 $\sim$ 1.05 Mcounts/spill for 
MS1, 2 = $\pm$2.5 mm.
The beam yield ratio between MS1, 2 = $\pm$2.5 mm and  $\pm$0.7 mm  is 
2.6 $\sim$ 3.0.
Therefore, 
the $\overline{p}$ beam yield for MS1, 2 = $\pm$2.5 mm is estimated to be 
$\sim$ 0.3 Mcounts/spill from measured yield of 95 kcounts/spill 
for MS1, 2 = $\pm$ 0.7 mm in RUN\#5390.

\begin{table}[h]
\centering
\caption{Slit, ESS and CM conditions in $K^-$ beam study shown in Fig.~\ref{fig-kvsslit}. 
}
\label{tab-kcond}       
\begin{tabular}{cccccccccc}
\hline
IFV  & IFH & MOM & MS1/MS2   & ESS1 & ESS2 & CM1/CM2 & CM3/CM4 \\
$[$mm$]$ & $[$mm$]$ & $[$mm$]$  
&  $[$mm$]$ & $[$kV$]$ & $[$kV$]$ & $[$A$]$ & $[$A$]$\\
\hline
$^{+1.3}_{-0.9}$  & $\pm$120 & $\pm$180  & --/-- 
& $\pm$250 &  $\pm$250  &  402/380 &  421/349  \\  
\hline
\end{tabular}

\end{table}

\section*{Appendix B ~Estimation of optimized $\overline{p}$ beam 
yield from $\overline{p}$/$K^-$ ratio} 

A systematic study of $\overline{p}$ beam and $K^-$ beam tuning was performed after 
the T1 target exchange period in 2019 as summarized in Table \ref{tab-e40beamstudy}.
Firstly, center values of CM1, 2 or CM3, 4 currents 
were scanned for $\overline{p}$ (A and B).
After that, the offset values 
of  CM1, 2 and CM3, 4 currents were scanned (C).
(e.g. The offset value is $\pm$10 A for CM1, 2 of 435  A/415 A in (C)
compared to the center value of 425 A in (B).) 
Then, the gain of
the  $\overline{p}$ beam yield with CM offset was found to be 1.2 (B $\to$ C).
The $K^-$ beam yield was obtained to be 170k for MS1, 2 = $\pm$0.5 mm as
shown in E.
Even we did not take a directly comparable data for $K^-$  and $\overline{p}$ beams  
with the same slit condition and CM offset at that time,
the $\overline{p}$ beam yield 
can be estimated to be 162k (A $\times$ 1.2) 
which is comparable with the $K^-$ beam yield of 170k (E).
%
%
%
In addition, 
a $\overline{p}/K^-$ beam yield comparison was made with the same slit condition 
in 2015 as shown in Table \ref{tab-e13beamstudy}.
Then, the $\overline{p}$ and $K^-$ beam yields were measured  to be 75k and 
62k, respectively.
From these results, we concluded that 
the $\overline{p}$/$K^-$ beam ratio is roughly $\sim$ 1.

On the other hand, 
using the Sangford-Wang parameterization \cite{SW} with kinematical reflection factor, 
the $\overline{p}$/$K^-$ yield ratio  is estimated to be 
0.7 which is in agreement with our conclusion based on measurements.

Since the $K^-$ beam yield for MS2 =  $\pm$0.7 mm 
was measured to be 350 kcounts/spill 
as shown in Fig.~\ref{fig-kvsslit} for well tuned condition,
the  $\overline{p}$ beam yield is also 
expected to be 245 $\sim$
350 kcounts/spill with the same condition
using the 
ratio of $0.7 \sim 1$.

\begin{table}[h]
\centering
\caption{1.8 GeV/$c$
$\overline{p}$ beam and $K^-$ beam yields for several MS/CM conditions 
for ESS1, 2 = $\pm$250 kV and IFV  opening width of 2.0 mm 
in 2019 (Acc.~RUN85).
MR power was 50  kW ($55 \times 10^{12} $ protons/spill), and 
the T1 target and repetition rate  were the same as 
in this $\overline{d}$ beam study (Acc.~RUN87).
$\overline{p}$ beam yield of D is estimated value using A--C.
}
\label{tab-e40beamstudy}       
\begin{tabular}{lcccc|ccc}
\hline
      &  MS1 & MS2                & CM1/CM2 & CM3/CM4  &$\overline{p}$  &  $K^-$\\
      &  $[$mm$]$& $[$mm$]$ & $[$A$]$ & $[$A$]$ & $[$counts/spill$]$ & 
$[$counts/spill$]$\\ 
\hline
A    &  $\pm 0.5$& $\pm 0.5$ &425  &415  & 135k \\
B    &  $\pm 0.5 $& $\pm 1.0$ &425 &415  & 264k \\  
C               &   $\pm 0.5 $& $\pm 1.0$ &435/415&444/384 & 318k \\
D       &   $\pm 0.5 $& $\pm 0.5$ &435/415&444/384 & (162  k) \\
\hline
E               &   $\pm 0.5$& $\pm 0.5$ &400/380 &415/351 & &170k \\
\hline

\end{tabular}
\end{table}

\begin{table}[h]
\centering
\caption{
1.8 GeV/$c$ $\overline{p}$ beam and $K^-$ beam yields for the same slit condition 
with  ESS1, 2 = $\pm$250 kV measured in 2015 (Acc.~RUN63).
MR power was 24  kW($33 \times 10^{12}$ protons/spill) with 
a different T1 target (Au 66 mm thickness)
and repetition rate (5.52 s).
IFV opening width was 2.0 mm.
}
\label{tab-e13beamstudy}       
\begin{tabular}{ccccc|ccc}
\hline
     &  MS1 & MS2                & CM1/CM2 & CM3/CM4  &$\overline{p}$  &  $K^-$\\ 
     &  $[$mm$]$& $[$mm$]$ & $[$A$]$ & $[$A$]$ & $[$counts/spill$]$ & 
$[$counts/spill$]$\\ 
\hline
       &  $\pm 0.5 $& $\pm 0.5$ &439&414 & 75k \\
       &  $\pm 0.5 $& $\pm 0.5$ &405&380 & &62k \\
\hline

\end{tabular}
\end{table}

\end{document}